**Sample size calculation comparing two survival curves in the presence of nonproportional hazards utilizing the concept of Weighted Relative Time using two different Weibull distributions**

**Authors**: Milind A. Phadnis[1], Md Robiul Islam Talukder[1], Geethanjalee Mudunkotuwa[1], Matthew S. Mayo[1]

**From**: [1]Department of Biostatistics & Data Science, University of Kansas Medical Center, Kansas City, KS, USA.

**Corresponding author:** Milind A. Phadnis PhD[1], 3901 Rainbow Boulevard, Kansas City, KS 66160; Phone 913-945-7986; FAX 913-588-7963; email mphadnis@kumc.edu

**Disclaimer:** None

**Disclosure:** The authors have no relevant conflicts of interest to declare.

R**unning Title:** Sample size for nonproportional hazards using Weighted Relative Time

**ABSTRACT:** Sample size calculations in two-arm clinical trials with survival endpoint often utilize the proportional hazards (PH) assumption. In case of nonproportional hazards (NPH), much research is ongoing with some newer methods developed that work well in specific scenarios. Recent developments have considered the nonparametric Max-Combo logrank test for both design and analysis of survival data with NPH. Although many approaches have been proposed for analyzing time-to-event data with NPH such as Restricted Mean Survival Time, Average Hazard Ratios, and piecewise exponential model with hazard ratios changing over time, translation into sample size calculation is not obvious and relies on simulations for very specific scenarios of NPH. In the parametric setting, accelerated failure time models that allow NPH but require a constant time ratio assumption (PT assumption) have been proposed. Recently, a parametric approach that simultaneously allows both nonproportional hazards (NPH) and nonproportional time (NPT) has been proposed. It utilizes the concept of Relative Time (RT) with two different Weibull distributions for the two trial arms in the case of monotonically increasing/decreasing time ratios. A key limitation is that the hypotheses leading to the sample size calculation rely on a single effect size defined at a specific survival percentile. In our current work, we extend the parent RT method by utilizing the concept of Weighted Relative Time (WRT) that allows us comparison of the full survival curves. Performance of the proposed method is compared to the Max-Combo method under different weighting schemes, and its operation characteristics are studied via simulations.

## 1. Introduction

Traditional sample size calculations for two-arm trials with a time-to-event endpoint are often conducted using the proportional hazards (PH) assumption, or by assuming that the survival time follows an exponential distribution. In this context, standard statistics software provides numerous options such as the Bernstein and Lagakos F-test[1] formula for exponential distributed survival times, nonparametric logrank test of Freedman[2], Lachin and Foulkes[3], and the Schoenfeld[4,5] formula for the semiparametric PH model of Cox[6]. The general approach is to first calculate the number of events required to have a prespecified power (say 80%) to detect a treatment effect of clinical interest keeping the type I error at a prespecified level (say 5%). The total sample size is then obtained by inflating the number of events to account for administrative censoring and random dropouts. For more advanced multi-arm trials with complex and adaptive features where no closed-form formula for calculating the sample size is readily available, a simulation-based approach can be adopted both in the frequentist and Bayesian setting. For example, the FACTS[7] software allows a statistician to study the operation characteristics (OC) of a simulated trial with many advanced (Bayesian and frequentist) adaptive features (assuming PH) with an initial estimate of sample size. If the OC are not satisfactory, the initial estimate can be updated and the process repeated till satisfactory OC are obtained. However, the performance of these methods depends on the validity of underlying assumptions. Thus, the Schoenfeld formula is built specifically to handle PH assumption, and the logrank test performs optimally only in the case of PH. Lakatos and Lan[8] discuss the performance of the logrank statistic in PH and NPH cases, and under the exponential distribution. Likewise, FACTS utilizes prespecified piecewise constant hazards for the control arm, however, the effect size is defined by means of a constant hazard ratio (PH assumption).

When analyzing survival data, the PH assumption can be assessed diagnostically and adjustments accounting for NPH can be made to draw meaningful inferences from the analyzed data. For example, one can posit that the hazard ratio (HR) changes over time and can adopt several methods that model this phenomenon – piecewise exponential models, extended Cox model with a *treatment arm by time* interaction, parametric accelerated failure time models that allow NPH but assume constant time ratio (PT) assumption, Restricted Mean Survival Time (RMST), and the Max-Combo test – to name a few. A key topic of research interest is the design of randomized clinical trials when the PH assumption is not valid (nonproportional hazards – abbreviated as NPH) and it is not easy to translate the above-mentioned analysis methods into calculating sample size as part of the design problem. One must often rely on extensive simulations and there is no single approach that is optimal for handling the myriad patterns of NPH for time-to-event endpoints. The review paper by Ananthakrishnan et al.[9] provides a summary of the different methods that can be used to design trials in the presence of specific scenarios of NPH along with the associated advantages and limitations.

A key conceptual idea in designing a trial with NPH is deciding the *most appropriate definition* of *effect size*. Even in the PH case, it is harder to convey the hypothesized benefit of a new treatment (compared to control) to potential participants through the language of *'percent instantaneous risk reduction'* than the more intuitive idea of *'improvement in median longevity'*. That is, from the point of view of the patients and their families, the idea that if treatment is shown to perform better than the control, the sentence *'median survival improved from 6 months to 12 months'* is more clearly interpretable than *'the instantaneous risk of death reduced by 25%'*. Under NPH, this problem gets even more compounded. Here, the NPH scenarios warrants a hazard ratio that is not constant over time – that is, for two different times $t_1 \neq t_2$, we may have $HR(t_1) \neq HR(t_2)$. Thus, it will be even harder to convey the hypothesized survival benefit through the language *"at an earlier time point $t_1$, the instantaneous risk reduction for death is (say) 25% and at a later time point $t_2$, the instantaneous risk reduction for death is (say) 40%'*. Instead, the language *"for early failures (say at the 25$^{th}$ percentile), longevity will improve from 4 months to 6 months, and for late failures (say at the 75$^{th}$ percentile), longevity will improve from 9 months to 18 months"* is more interpretable. Motivated by these considerations, Phadnis and Mayo[10] have advocated the metric of Relative Time denoted by RT(p) – time-ratio that changes over percentile *p* – as a measure of the effect size in designing clinical trials. In its most general form, RT(p) simultaneously allows both NPH and NPT. The Weibull AFT model is a special case that simultaneously allows both PH and PT, while non-Weibull AFT models are special cases that allow NPH but with PT. The RT(p) concept is discussed in detail in Cox et al.[11] and critical ideas with notation are reproduced in the Appendix (A.1 – A.3). Phadnis and Mayo[10] have developed a test statistic that asymptotically follows a normal distribution allowing them to first calculate the number of events, and then to calculate the total sample size accounting for administrative censoring for fixed two-arm trials. Sharma and Phadnis[12,13] have utilized this concept for stochastic curtailment tests with NPH, and Phadnis et al.[14] has extended it to group sequential testing. More recently, Thewarapperuma and Phadnis[15] have extended it to multi-arm trials with adaptive features in a Bayesian setting. Despite these promising developments, the one major limitation of the Phadnis and Mayo[10] is that the hypotheses leading to the sample size calculations is of the type $\mathrm{H}_0$: $RT(p_{mid}) \leq 1$ versus $\mathrm{H}_1$: $RT(p_{mid}) > 1$ (or its two-sided counterpart). That is, even though the method handles NPH, the calculated sample size maintains the target power to detect an effect size of magnitude $RT(p_{mid})$ – *time ratio at any convenient user-defined percentile of survival time* – with a prespecified type I error. In the current work, we extend our previous work by introducing the concept of Weighted Relative Time (explained later) to compare the two full survival curves in the general NPH scenario.

## 2. Motivating examples

Two motivating examples discussed in Phadnis and Mayo[10] are illustrated here to highlight the areas of application for the proposed method.

The first example is for a phase III two-arm study for patients undergoing an initial treatment followed by a second-line treatment for chemotherapy refractory advanced metastatic cholangiocarcinoma which is a rare aggressive neoplasm. The endpoint of interest is progression-free survival (PFS) and researchers hypothesize that a new experimental treatment (E) will improve the PFS compared to a standard control (C). The C arm has been well studied in previous phase II trials where the median PFS has been reported as 4 months with an interquartile range of 2-7 months. Due to the small value of median PFS, the researchers wish to express their hypotheses in the metric of time and thereby be able to convey to the patients the hypothesized PFS benefit in terms of *improved longevity.* They hypothesize that the *improvement in longevity* for the $10^{th}$ percentile of PFS will be by a factor of 1.5 and that the treatment will show a gradual monotonic trend with the *improvement in longevity* for the $90^{th}$ percentile of PFS will be by a factor of 2. Thus, this scenario reflects the researcher's hypothesized benefit is not only for increased median PFS, but that it gradually improves over time. Accrual for the study occurs at a uniform rate over the enrollment period of 12 months. The study follow-up time is 12 months; target power is 80% and type I error is fixed at 5% for a one-sided hypothesis (which is acceptable for rare cancers). This scenario is best elicited through Figure 1 and represents frequently occurring situations in oncology trials where researchers expect long-term survivors to receive maximum benefit from a new treatment with only small realistic improvements expected of short-terms survivors.

The second example (illustrated by Figure 2) explores a different scenario of a two-arm trial (surgery vs non-surgical standard-of-care control) in which the outcome of interest is Overall Survival (OS). Here, researchers hypothesize that patients receiving surgery will experience an immediate significant survival benefit, but that this improvement will wane over time. While the researchers anticipate a significant survival benefit in terms of median OS, they hypothesize that for the 10$^{th}$ percentile of OS will be by a factor of 2 and the *improvement in longevity* for the 90$^{th}$ percentile of OS will be by a factor of 1.5.

In both examples, trialists collaborating with the statistician would like to perform a sensitivity analysis by varying different design parameters and are open to considering somewhat larger values of accrual and follow-up time.

## 3. Methods

The utilization of the RT(p) concept for sample size calculation is discussed in detail in Phadnis and Mayo[10] and critical ideas with the mathematical notation are reproduced in the Appendix (A.1 – A.3). Herein, we focus on extending the method of Phadnis and Mayo[10] to allow sample size calculation for comparing two survival curves (from two different Weibull distributions) in the presence of nonproportional hazards. In doing so, we keep in mind the key user inputs $p_I, RT(p_I), p_{II},$ and $RT(p_{II})$ mentioned in Appendix A.1.

### 3.1. *Setting up the hypotheses*

Consider any convenient choice of $k$ percentiles $p_1 < p_2 < p_3 < \cdots < p_k$ of survival time. For example, $k = 2$ could be a choice with $p_1 = 0.1$ and $p_2 = 0.9$. Another choice could be $k = 3$ with $p_1 = 0.25, p_2 = 0.5$, and $p_3 = 0.75$. Other possibilities are $k = 9$ with $p_1 = 0.1, p_2 = 0.2, p_3 = 0.3, \ldots, p_9 = 0.9$, or even $k = 99$ with $p_1 = 0.01, p_2 = 0.02, p_3 = 0.03, \ldots, p_{99} = 0.99$. As shown later, these different choices of $k$ can be a part of *sensitivity analysis* for the proposed sample size calculation. Note also that $p_I \leq p_1$ and likewise $p_{II} \geq p_k$. Now, instead of the one-sided hypotheses $\mathrm{H}_0$: $RT(p_{mid}) \leq 1$ versus $\mathrm{H}_1$: $RT(p_{mid}) > 1$ or its two-sided counterpart $\mathrm{H}_0$: $RT(p_{mid}) = 1$ versus $\mathrm{H}_1$: $RT(p_{mid}) \neq 1$ given in Phadnis and Mayo[10], we consider the more general hypotheses given below:

One-sided test $\forall\, p$:

$$\mathrm{H}_0:\ RT(p) \leq 1$$
$$H_1:\ RT(p) > 1 \tag{1}$$

Two-sided test $\forall\, p$:

$$\mathrm{H}_0:\ RT(p) = 1$$
$$H_1:\ RT(p) \neq 1 \tag{2}$$

Before proceeding further, it is important to recall the notation of $q_{min}$ and $q_{max}$ mentioned in Section 3.6 of Phadnis and Mayo[10]. Consider the survival time in arm $i$ ($i = 0, i = 1$ for control and treatment arm respectively) follows the Weibull distribution $T_i \sim Weibull(\beta_i, \theta_i)$ with shape parameter $\beta_i$ and scale parameter $\theta_i$. Then, $q_{min}$ is the smallest value for $p$ at which the crossing of two curves is permitted as considered plausible based on biological/clinical considerations. Thus $0 < q_{min} < p_1 < p_2 < \cdots < p_k$ gives the range for $q_{min}$ and can play a role in the sample size calculation when $RT(p_1) < RT(p_2) < \cdots < RT(p_k)$. Analogously, $q_{max}$ is the largest value for $p$ at which the crossing of two curves is permitted.

Thus $p_1 < p_2 < \cdots p_k < q_{max} < 1$ gives the range for $q_{max}$ and plays a role in the sample size calculation when $RT(p_k) > \cdots > RT(p_2) > RT(p_1)$. As a simple example, if the collaborator guided inputs are: $k = 2,\ p_I = p_1 = 0.1, p_{II} = p_2 = 0.9, RT(p_1) = 1.5, RT(p_2) = 2$, then this combination results in crossing of the two survival curves at $q_{min} = 0.00135, RT(q_{min}) = 1 \Rightarrow \forall\, q < q_{min}, RT(q) < 1$ and this means that for all very small percentiles (very early in the observation window), survival in C is higher than survival in E. In most situations, from a practical standpoint, such small values of $q_{min}$ will be clinically acceptable. Phadnis and Mayo[10] discuss some corrective measures when this is not true.

Thus, more accurately the two-sided hypotheses in (2) can be restated as:

Two-sided test where $RT(p_1) < RT(p_2) < \cdots < RT(p_k)\ \forall\, p > q_{min}$:

$$\mathrm{H}_0\text{:}\ \ RT(p) = 1$$

$$H_1\text{:}\ \ RT(p) \neq 1 \tag{3}$$

or

Two-sided test where $RT(p_1) > RT(p_2) > \cdots > RT(p_k)\ \forall\, p < q_{max}$:

$$\mathrm{H}_0\text{:}\ \ RT(p) = 1$$

$$H_1\text{:}\ \ RT(p) \neq 1 \tag{4}$$

with corresponding changes for the one-sided hypotheses in (1) given by $\mathrm{H}_0\text{: } RT(p) \leq 1$ versus $\mathrm{H}_1\text{:}\ \ RT(p) > 1$.

Recall the null hypotheses for a weighted logrank test in standard statistical texts such as Klein and Moeschberger[16] given as $\mathrm{H}_0\text{:} h_1(t) = h_2(t)\ \ \forall\, t \leq t_{max}$ or in Allison[17] as $\mathrm{H}_0\text{:} S_1(t) = S_2(t)\ \ \forall\, t \leq t_{max}$ where $t_{max}$ is the largest event time and the subscripts 1 and 2 index the two groups C and E. This could be equivalently written as $\mathrm{H}_0\text{:} HR(t) = 1\ \ \forall\, t \leq t_{max}$ with $HR(t) = \frac{h_1(t)}{h_2(t)}$ and doing so allows us to compare it to the hypotheses in (3) and (4). While the hypotheses based on the weighted logrank test adopt a hazard ratio metric applicable to all event times less than or equal to $t_{max}$, our hypotheses given in (3) and (4) adopt a *Relative Time* metric applicable to all survival percentiles greater than $q_{min}$ when $RT(p_1) < RT(p_2) < \cdots < RT(p_k)$, or, less than $q_{max}$ when $RT(p_1) > RT(p_2) > \cdots > RT(p_k)$.

3.2. *Formulation of a test statistic*

Consider that the survival percentiles $p_j$ for $j = 1,2, \dots, k$ with $k \geq 2$ have been specified with corresponding treatment effect sizes $RT(p_j)$.

Our statistic of interest is:

$$Q_w = \sum_{j=1}^{k} w_j(p_j) \log\{RT(p_j)\} \tag{5}$$

where $\sum_{j=1}^{k} w_j(p_j) = 1$ and $w_j(p_j)$ is the weight at $p_j$. The subscript $w$ on $Q_w$ is only written to differentiate it from $Q'$ used in Phadnis and Mayo[10]. We can also write the variance of $Q_w$ as:

$$\mathrm{Var}(Q_w) = \sum_{j=1}^{k} w_j(p_j)^2 \, \mathrm{Var}[\log\{RT(p_j)\}]$$

$$+ 2 \sum_{j=1}^{k} \sum_{l>j}^{k} w_j(p_j) w_l(p_l) \, \mathrm{Cov}[\log\{RT(p_j)\}, \log\{RT(p_l)\}] \tag{6}$$

Recall that in Phadnis and Mayo[10], the test statistic $Q' = \log(Q_1/Q_0) \sim \mathrm{N}(\mu_d, \sigma_d^2)$ with $\mu_d = \log\left[\frac{\theta_1}{\theta_0}\left\{\log\left(\frac{1}{1-p_{mid}}\right)\right\}^{\frac{1}{\beta_1}-\frac{1}{\beta_0}}\right]$ and $\sigma_d = \sqrt{\frac{1}{d_1\beta_1^2} + \frac{1}{d_0\beta_0^2}}$ where $d_0$ and $d_1$ are the number of events in the control arm and treatment arm respectively and $r$ is the allocation ratio. This allowed them to calculate $d_0$ and $d_1$ using the fact that $Z = \frac{Q'-\mu_d}{\sigma_d}$ is distributed $\mathrm{N}(0,1)$ under $\mathrm{H}_0$. In our context, since we have $p_j$ for $j = 1,2, \dots, k$ with $k \geq 2$, we can write $Q'_j = \log(Q_{1j}/Q_{0j}) \sim \mathrm{N}\left(\mu_{d_j}, \sigma^2_{d_j}\right)$ with $\mu_{d_j} = \log\left[\frac{\theta_1}{\theta_0}\left\{\log\left(\frac{1}{1-p_j}\right)\right\}^{\frac{1}{\beta_1}-\frac{1}{\beta_0}}\right]$ and

$\mathrm{Var}\left[\log\left\{RT\left(p_j\right)\right\}\right] = \sigma^2_{d_j} = \frac{1}{d_{1j}\beta_1^2} + \frac{1}{d_{0j}\beta_0^2}$ where $d_{0j}$ and $d_{1j}$ are the number of events in the control arm and treatment arm respectively corresponding to the time-ratio of $RT\left(p_j\right)$ at $p_j$.

Due to normality of $Q'_j$, we have $E(Q_w) = \mu_w = \sum_{j=1}^{k} w_j\left(p_j\right)\mu_{d_j}$ and under the null hypotheses of (3) or (4), we have $Z_w = \frac{Q_w-\mu_w}{\sqrt{\mathrm{Var}(Q_w)}} \sim \mathrm{N}(0,1)$.

Thus, the number of events in the control arm denoted by $d_{0_w}$ and the number of events in the treatment arm denoted by $d_{1_w}$ required to detect a treatment effect of size $Q_w$ (under the alternate hypotheses) with $100\omega\%$ power with a prespecified type I error $\alpha$ is given by:

For a one-sided hypotheses:

$$d_{0_w} = \left[\frac{(Z_\omega + Z_{1-\alpha})}{Q_w}\right]^2\left(\frac{1}{r_w\beta_1^2} + \frac{1}{\beta_0^2}\right)$$

$$d_{1_w} = r_w d_{0_w} \tag{7}$$

For a two-sided hypotheses:

$$d_{0_w} = \left[\frac{\left(Z_\omega + Z_{1-\alpha/2}\right)}{Q_w}\right]^2\left(\frac{1}{r_w\beta_1^2} + \frac{1}{\beta_0^2}\right)$$

$$d_{1_w} = r_w d_{0_w} \tag{8}$$

where $r_w = d_{1_w}/d_{0_w}$ is the allocation ratio.

Next, consider a trial with $d_{0_w}$ and $d_{1_w}$ events in the control and treatment arm, with fixed $\beta_0$ and $\beta_1$, and prespecified values for any pair of survival percentiles $p_j, p_l$ (where $j = 1,2,\ldots,k$ and $l = 1,2,\ldots,k$ and $l > j$ with $k \geq 2$) with corresponding pairs of treatment effect sizes $\mathrm{RT}(p_j)$, $\mathrm{RT}(p_l)$. Then the theoretical correlation $\rho_{ij} = \mathrm{Corr}\left[\log\{RT(p_j)\}, \log\{RT(p_l)\}\right] = 1$ due to $p_j < p_l$. Also as shown in Appendix A.4, the empirical correlation $\hat{\rho}_{ij} =$

$\text{Corr}\left[\log\{\widehat{RT}(p_j)\}, \log\{\widehat{RT}(p_l)\}\right] = 1$ due to the fact that $p_j < p_l$ (by definition),

$\log\{\widehat{\text{RT}}(p_j)\} \dot{\sim} \text{N}\left(\log\left[\frac{\hat{\theta}_1}{\hat{\theta}_0}\left\{\log\left(\frac{1}{1-p_j}\right)\right\}^{\frac{1}{\beta_1}-\frac{1}{\beta_0}}\right], \frac{1}{d_{1_w}\beta_1^2} + \frac{1}{d_{0_w}\beta_0^2}\right)$ and

$\log\{\widehat{\text{RT}}(p_l)\} \dot{\sim} \text{N}\left(\log\left[\frac{\hat{\theta}_1}{\hat{\theta}_0}\left\{\log\left(\frac{1}{1-p_l}\right)\right\}^{\frac{1}{\beta_1}-\frac{1}{\beta_0}}\right], \frac{1}{d_{1_w}\beta_1^2} + \frac{1}{d_{0_w}\beta_0^2}\right)$. That is, $\log\{\widehat{\text{RT}}(p_j)\}$ and

$\log\{\widehat{\text{RT}}(p_l)\}$ are both normally distributed with the same variance. In the scenario (of Figure 1)

where $\text{RT}(p_j) < \text{RT}(p_l)$, we have $\log\left[\frac{\hat{\theta}_1}{\hat{\theta}_0}\left\{\log\left(\frac{1}{1-p_j}\right)\right\}^{\frac{1}{\beta_1}-\frac{1}{\beta_0}}\right] < \log\left[\frac{\hat{\theta}_1}{\hat{\theta}_0}\left\{\log\left(\frac{1}{1-p_l}\right)\right\}^{\frac{1}{\beta_1}-\frac{1}{\beta_0}}\right]$. In the

scenario (of Figure 2) where $\text{RT}(p_j) > \text{RT}(p_l)$, we have $\log\left[\frac{\hat{\theta}_1}{\hat{\theta}_0}\left\{\log\left(\frac{1}{1-p_j}\right)\right\}^{\frac{1}{\beta_1}-\frac{1}{\beta_0}}\right] >$

$\log\left[\frac{\hat{\theta}_1}{\hat{\theta}_0}\left\{\log\left(\frac{1}{1-p_l}\right)\right\}^{\frac{1}{\beta_1}-\frac{1}{\beta_0}}\right]$. This can be verified through simulations.

Thus, the term $\text{Cov}\left[\log\{RT(p_j)\}, \log\{RT(p_l)\}\right]$ in (6) can be approximated as:

$$\text{Cov}\left[\log\{\text{RT}(p_j)\}, \log\{\text{RT}(p_l)\}\right] \approx \sqrt{\text{Var}\left[\log\{\text{RT}(p_j)\}\right]\text{Var}\left[\log\{\text{RT}(p_l)\}\right]} \quad (9)$$

Then assuming uniform accrual with accrual time $a$ and follow-up time $f$, event rates in the control and treatment arm accounting for administrative censoring can be calculated using the formula for $v_i, i = 0,1$ given in equations A.3.2 and A.3.3 of the Appendix.

Then with $v = (v_0 + v_1)/2$, sample size $n_{0_w} = d_{0_w}/v$ in the control arm and sample size $n_{1_w} = d_{1_w}/v$ in the treatment arm accounting for administrative censoring can be calculated.

The entire process can be summarized in the following steps.

Step 1: Elicit information about trial inputs (after discussion with collaborators) – $\alpha, \omega, p_I, p_{II}, \text{RT}(p_I)$, and $\text{RT}(p_{II})$ to represent the scenario in Figure 1 or Figure 2.

Step 2: Choose a weighting scheme $w_j(p_j)$ for for $j = 1,2, \dots, k \geq 2$ based on clinical considerations (*relative importance of early, middle and late treatment effects)* and the desired allocation ratio $r_w$. Use properties of the Weibull distributions to calculate $\mathrm{RT}(p_j)$.

Step 3: Calculate $d_{0j}$ and $d_{1j}$

Step 4: Calculate $d_{0_w}$ and $d_{1_w}$ using equations (5) – (9)

Step 5: Calculate $n_{0_w}$ and $n_{1_w}$ to account for administrative censoring. Inflate further by 1 minus the anticipated dropout rate to account for random loss to follow-up.

3.3. *Discussion of weighting schemes generated by* $w_j(p_j)$

Recall that the numerator of the test statistic for a weighted logrank test for the hypotheses $\mathrm{H}_0: S_1(t) = S_2(t) \;\; \forall\, t \leq t_{max}$ is given by $A = \sum_{j=1}^{k} w_j(t_j)[O_j - E_j]$ where $O_j$ and $E_j$ are observed and expected number of events at event times $t_j$. Thus, the weights $w_j(t_j)$ drive the relative importance given to early, middle or late failures. A common approach is to use the weights given by Fleming and Harrington[18] of the type $w_{a,b}(t_j) = \hat{S}(t_{j-1})^a[1 - \hat{S}(t_{j-1})]^b$ such that the combination $a = b = 0$ yields the logrank test (with all failure times weighed equally). Likewise, $a > 0, b = 0$ gives more weight to early failures, and $b > 0, a = 0$ gives more weight to late failures. In our context too, we adopt a similar approach except that our weights are of the type $w_j(p_j)$ instead of $w_j(t_j)$.

Consider the case with $k = 9$ with $p_1 = 0.1, p_2 = 0.2, p_3 = 0.3, \dots, p_8 = 0.8, p_9 = 0.9$. User defined raw scores $u_j(p_j) > 0$ can be applied to the survival percentiles $p_j$ with $w_j(p_j) = \frac{u_j(p_j)}{\sum_{j=1}^{k} u_j(p_j)}$ such that $\sum_{j=1}^{k} w_j(p_j) = 1$ to represent various scenarios of giving more or less importance to early, middle or late failures. Thus, the raw scores $u_j(p_j) \equiv 0,0,0,0,1,0,0,0,0$ would represent the special case of sample size calculations done when comparing median survival times of the two treatment arms. Likewise, $u_j(p_j) \equiv 1,1,1,1,1,1,1,1,1$ represents the scenario with equally weighted failure times, $u_j(p_j) \equiv 9,8,7,6,5,4,3,2,1$ represents more weight to early times with weights reducing over time and $u_j(p_j) \equiv 1,2,3,4,5,6,7,8,9$ represents the exact opposite. Another example would be $u_j(p_j) \equiv 0,0,0.5,1,2,1,0.5,0,0$ with only middle failures considered to be important in conducting the sample size calculations. Infinite flexible weighting patterns can be created, and many such patterns are discussed in detail in the Results section. Our method can thus be called the *Weighted Relative Time (WRT)* method.

### 3.4 *The special case of test of medians*

A direct advantage of our proposed WRT method is that any weighting scheme with $u_j(p_j) = 1$ for $p_j = 0.5$ and $u_j(p_j) = 0$ for $p_j \neq 0.5$ becomes a test of medians of two Weibull distributions. That is, the one-sided hypotheses $H_0$: $RT(p_{median}) \leq 1$ versus $H_1$: $RT(p_{median}) > 1$ or its two-sided counterpart $H_0$: $RT(p_{median}) = 1$ versus $H_1$: $RT(p_{median}) \neq 1$ can be tested accounting for non-proportionality of hazards owing to the knowledge of $\beta_0$ and $\beta_1$ in the sample size formula (see the weighting scheme WS08 in Section 4.1). It is quite common for researchers to ask a statistician to design a trial with effect size defined purely through median survival times of the control and treatment arm. Then the two common approaches used by statisticians are: {i} exponentially (parametric) distributed survival times – which can be seen as a special case of our WRT method with the additional restriction of $\beta_0 = \beta_1 = 1, RT(p_I) = RT(p_{II})$ , and, separately {ii} PH assumption (semiparametric) which can be seen as a special case of our WRT method with the additional restriction of $\beta_0 = \beta_1, RT(p_I) = RT(p_{II})$. It should be noted that the Brookmeyer-Crowley test of medians[21] is a popular nonparametric method to conduct a hypotheses test of medians for time-to-event endpoint, but at the design stage it could only be used through extensive simulations assuming specific shapes for the survival curves of the two arms. Thus, our proposed WRT method provides a solid alternate option for designing two-arm trials based on improvement in median longevity whilst accounting for non-PH and non-PT scenarios.

### 3.5 *The special case of equal weights and its connection to test of means of two arms*

Section 3.3 mentioned the example of $k = 9$ with $p_1 = 0.1, p_2 = 0.2, \dots, p_8 = 0.8, p_9 = 0.9$ with equal raw scores $u_j(p_j) \equiv 1,1,1,1,1,1,1,1,1$ leading to equal weights $w_j(p_j) \equiv \frac{1}{9}, \frac{1}{9}, \dots, \frac{1}{9}, \frac{1}{9}$. Although other values of $k$ are discussed in Section 4.3 (sensitivity analyses), here we bring attention to the scenario when $k$ is very large. As an example, $k = 99$ would allow the equal weights $w_j(p_j) \equiv \frac{1}{99}, \frac{1}{99}, \dots, \frac{1}{99}, \frac{1}{99}$ with $p_1 = 0.01, p_2 = 0.02, \dots, p_{98} = 0.98, p_{99} = 0.99$. Likewise, $k = 999$ would allow the equal weights $w_j(p_j) \equiv \frac{1}{999}, \frac{1}{999}, \dots, \frac{1}{999}, \frac{1}{999}$ with $p_1 = 0.001, p_2 = 0.002, \dots, p_{998} = 0.998, p_{999} = 0.999$. For a given $k$, the test statistic $Q_w$ from equation (5), would simplify to:

$$Q_w = \frac{1}{k}\sum_{j=1}^{k} \log\{RT(p_j)\} = \frac{1}{k}\sum_{j=1}^{k} \log\left\{\frac{S_1{}^{-1}(1-p_j)}{S_0{}^{-1}(1-p_j)}\right\}$$
$$= \log\left[\left\{\prod_{j=1}^{k} t_1(p_j)\right\}^{\frac{1}{k}}\right] - \log\left[\left\{\prod_{j=1}^{k} t_0(p_j)\right\}^{\frac{1}{k}}\right] = \log\left[\frac{\{\prod_{j=1}^{k} t_1(p_j)\}^{\frac{1}{k}}}{\{\prod_{j=1}^{k} t_0(p_j)\}^{\frac{1}{k}}}\right] \quad (10)$$

That is, $Q_w$ would be the logarithm of the ratio of the geometric means of survival times in the treatment arm $t_1(p_j)$ versus the control arm $t_0(p_j)$. Let $\Delta_p$ be the increment between successive equally spaced $p_j$ values. Then as $\Delta_p \rightarrow 0$, $k$ can be obtained as $k = 1 + \frac{p_{II} - p_I}{\Delta_p}$ with $p_I > q_{min}$ when $RT(p_I) < RT(p_{II})$ and $p_{II} < q_{max}$ when $RT(p_I) > RT(p_{II})$. As a result, $Q_w$ could be interpreted as the logarithm of the ratio of the geometric means of *"all possible"* survival times (events or censored) and these *"all possible"* times would span the area under the treatment and control survival curves contained between $p_I$ and $p_{II}$.

An approach that is gaining popularity for analyzing time-to-event data with nonproportional hazards is the *Restricted Mean Survival Time (RMST)* method of Royston and Parmar[22] where the area under the survival curves of the two study arms *(mean)* can be compared up to any restricted event time of interest $\tau$ and the maximum value for $\tau$ is the smaller of the largest event times in the two study arms. The ratio of the RMST of the two arms has been recommended for reporting by Trinquart et al.[23] in randomized controlled trials in oncology in presence of nonproportional hazards due to *ease of interpretation* for nonstatistical audiences. That is, in this case of RMST, the statistic of interest is a ratio of weighted arithmetic means restricted by $\tau$ – see expression for $\widehat{RMST}(\tau)$ in Royston and Parmar[22]. In contrast, our test statistic $Q_w$ is the logarithm of weighted geometric means restricted to lie between $p_I$ and $p_{II}$.

## 4. Results

### 4.1 *Sample size calculation using different weighting schemes*

Here, we discuss the cholangiocarcinoma trial (first example from Section 2) with the following design specifications: $\alpha = 0.05$ (type I error, one-sided), $100\omega = 80\%$ (power), $t_{med_0} = 4$ months (median survival time in the control arm), $a = 12$ months (accrual time with uniform accrual), $f = 12$ months (follow-up time), $r = 1$ (allocation ratio), $p_I = 0.1, \mathrm{RT}(p_I) = 1.5, p_{II} = 0.9, \mathrm{RT}(p_{II}) = 2$. To assess the sensitivity of the calculations, we have also assessed three alternate scenarios with $\mathrm{RT}(p_I) = 1.25$ and $\mathrm{RT}(p_{II}) = 1.75$, $\mathrm{RT}(p_I) = 2$ and $\mathrm{RT}(p_{II}) =$ 1.5, $\mathrm{RT}(p_I) = 1.75$ and $\mathrm{RT}(p_{II}) = 1.25$. Also, Figure 3 shows the different shapes of the hazard function generated by different values of the control arm shape parameter $\beta_0$ and treatment arm shape parameter $\beta_1$ for two different settings – {i} $p_I = 0.1, \mathrm{RT}(p_I) = 1.5, p_{II} = 0.9, \mathrm{RT}(p_{II}) = 2$ in Panel A, and {ii} $p_I = 0.1, \mathrm{RT}(p_I) = 2.0, p_{II} = 0.9, \mathrm{RT}(p_{II}) = 1.5$ in Panel B – reflecting different types of NPH scenarios.

We also consider 13 different weighting schemes that differentially weigh early, middle or late failures expressed through $u_j(p_j)$ for $j = 1,2, \ldots ,9$:

WS01 (Only very early effect important): $u_j(p_j) \equiv 1, 0, 0, 0, 0, 0, 0, 0, 0$
WS02 (Early effect important): $u_j(p_j) \equiv 1, 1, 1, 0, 0, 0, 0, 0, 0$
WS03 (Early and middle effect important): $u_j(p_j) \equiv 1, 1, 1, 1, 1, 1, 0, 0, 0$
WS04 (Sharp decreasing effect): $u_j(p_j) \equiv 9, 8, 7, 6, 5, 4, 3, 2, 1$
WS05 (Very gradual decreasing effect): $u_j(p_j) \equiv 3, 3, 3, 2, 2, 2, 1, 1, 1$
WS06 (Equal weight to all percentiles): $u_j(p_j) \equiv 1, 1, 1, 1, 1, 1, 1, 1, 1$
WS07 (Middle effect important): $u_j(p_j) \equiv 0, 0, 0, 1, 1, 1, 0, 0, 0$
WS08 (Only effect at median is important): $u_j(p_j) \equiv 0, 0, 0, 0, 1, 0, 0, 0, 0$
WS09 (Very gradual increasing effect): $u_j(p_j) \equiv 1, 1, 1, 2, 2, 2, 3, 3, 3$
WS10 (Sharp increasing effect): $u_j(p_j) \equiv 1, 2, 3, 4, 5, 6, 7, 8, 9$
WS11 (Late and middle effect important): $u_j(p_j) \equiv 0, 0, 0, 1, 1, 1, 1, 1, 1$
WS12 (Late effect important): $u_j(p_j) \equiv 0, 0, 0, 0, 0, 0, 1, 1, 1$
WS13 (Only very late effect important): $u_j(p_j) \equiv 0, 0, 0, 0, 0, 0, 0, 0, 1$

We call this the *Reference set* (other sets are considered in Section 4.3).

Table 1 displays the number of events (and sample sizes in parentheses) for each arm using equal allocation for different values of $\beta_0$ (shape parameter of the control arm) and the different weighting schemes mentioned above. For a given value of $\beta_0$ (say 0.5), when $\mathrm{RT}(p_I) = 1.5$ and $\mathrm{RT}(p_{II}) = 2$, the sample size expectedly decreases as we move from WS01 (only very early effect important) to WS06 (equal weight to all percentiles) to WS13 (only very late effect important). This is due to WS01 placing a weight of 1 on the smallest effect size of $\mathrm{RT}(p_1) = 1.5$ whereas WS13 places a weight of 1 on the largest effect size of $\mathrm{RT}(p_9) = 2$ with WS06 representing an intermediate option that weighs all $\text{RT}(p_j)$ for $j = 1,2, \dots 9$ equally with a weight of 1/9. A similar trend is seen when $\text{RT}(p_I) = 1.25$ and $\text{RT}(p_{II}) = 1.75$ but with corresponding sample sizes for each weighting scheme being larger compared to when $\text{RT}(p_I) = 1.5$ and $\text{RT}(p_{II}) = 2$. For a given value of $\beta_0$ (say 0.5), when $\text{RT}(p_I) = 2$ and $\text{RT}(p_{II}) = 1.5$, the sample size expectedly increases as we move from WS01 (only very early effect important) to WS06 (equal weight to all percentiles) to WS13 (only very late effect important). This is due to WS01 placing a weight of 1 on the largest effect size of $\text{RT}(p_1) = 2$ whereas WS13 places a weight of 1 on the smallest effect size of $\text{RT}(p_9) = 1.5$ with WS06 representing an intermediate option that weighs all $\text{RT}(p_j)$ for $j = 1,2, \dots 9$ equally with a weight of 1/9. A similar trend is seen when $\text{RT}(p_I) = 1.75$ and $\text{RT}(p_{II}) = 1.25$ but with corresponding sample sizes for each weighting scheme being larger compared to when $\text{RT}(p_I) = 2$ and $\text{RT}(p_{II}) = 1.5$. Similar expected trends are seen for $\beta_0 = 0.75, 1, \text{and } 1.25$.

For a given value of $\text{RT}(p_I)$ and $\text{RT}(p_{II})$ and a chosen weighting scheme, sample size decreases as we move from $\beta_0 = 0.5$ (very gradually decreasing Weibull hazard) to $\beta_0 = 0.75$ (decreasing Weibull hazard) to $\beta_0 = 1$ (exponential distribution with constant hazard) to $\beta_0 = 1.25$ (increasing Weibull hazard). This is again expected due to the property of the Weibull distribution as discussed in Wu[19] and in Phadnis[20]. Overall, our calculation of the number of events and sample sizes are as expected – more weights to percentiles representing smaller effect leads to larger sample size and vice versa. Additionally, for all scenarios, the special case WS08 represented by $u_j(p_j) \equiv 0, 0, 0, 0, 1, 0, 0, 0, 0$ yields the exact same answer as obtained by Phadnis & Mayo (2021) with sample size calculation conducted to have 80% power to detect an effect size represented by $\text{RT}(p_{mid})$ with the default value of $p_{mid}$ taken as 0.5. Likewise, Phadnis & Mayo (2021) discuss the possibility of calculating sample size at user-defined percentiles and these can also be obtained as special cases of our proposed method. That is, as an example, the sample size obtained by Phadnis and Mayo[10] for $\mathrm{RT}(p_{user-defined})$ with $p_{user-defined} = 0.3$ can be replicated by deploying a weighting scheme with $u_j(p_j) \equiv 0, 0, 1, 0, 0, 0, 0, 0, 0$. Thus, the method of Phadnis and Mayo[10] can now be seen as a special case of our current proposed method with exactly one $u_j(p_j) = 1$ and all other $u_j(p_j) = 0$. It should also be noted that the very special case of $\mathrm{RT}(p_I) = \mathrm{RT}(p_{II})$ necessarily leads to the proportional hazards assumption with $\beta_0 = \beta_1 = \beta$ and hence in this case our calculation reduces to the well-known Schoenfeld formula for calculating sample size in the PH scenario.

### 4.2 *Operation characteristics of our WRT method and comparison to the Max-Combo test*

The left panel of Table 2a displays the operation characteristics for our proposed WRT method. Taking $a$ = 12, f = 12, $p_I = 0.10$, $p_{II} = 0.90$ for various weighting schemes, effect sizes defined through $\mathrm{RT}(p_I)$ and $\mathrm{RT}(p_{II})$, and values of the control arm shape parameter $\beta_0 \equiv (0.5, 0.75, 1.0, 1.25)$, we conducted 10,000 simulations with a target type I error of 5% and a target power of 80% assuming 1:1 randomization for the two study arms. For the d/n (number of events/sample size) resulting from three weighting schemes (WS04, WS06, and WS10), we assessed empirical type I error, empirical power, average relative bias (ARB), mean square error (MSE) and coverage. For all scenarios, we found that the empirical type I error was very close to the target value of 0.05. Likewise, the empirical power was within the simulation error of the target power ($0.8 \pm 0.0078$). ARB and MSE were very low and coverage was always very close to the expected value of 90%. All simulations related to our WRT method were conducted using the SAS statistical software[24].

The right panel of Table 2a shows the operation characteristics – empirical type I error and empirical power obtained through 10,000 simulations of different settings of the MaxCombo test when applied to data consistent with our proposed WRT method. The MaxCombo test developed by Prior[25] is a generalization of the weighted log-rank test and is one of the most popular methods for analyzing time-to-event data in the presence of non-proportional hazards. Shen et al.[26] have discussed about its widespread and recommended use in cancer clinical trials and have expressed some concerns about the interpretability of results in certain non-PH scenarios such as *early separation of survival curves*. We considered 4 different sets of the MaxCombo – {a} Broad 6 set consisting of Fleming-Harrington component weights FH(0,0), FH(1,0), FH(0,1), FH(1,1), FH(2,0) and FH(0,2), {b} Classic 6 set consisting of Fleming-Harrington component weights FH(0,0), FH(1,0), FH(0,1) and FH(1,1) , {c} Early 4 set consisting of Fleming-Harrington component weights FH(0,0), FH(0.5,0), FH(1,0) and FH(2,0), and {d} Late 4 set consisting of Fleming-Harrington component weights FH(0,0), FH(0,0.5), FH(0,1) and FH(0,2). It should be noted that though we assessed operation characteristics of all 4 sets, the direct comparisons of interest between the WRT method and the MaxCombo are – {i} WS04 versus Early 4, {ii} WS06 versus Broad 6, {iii} WS06 versus Classic 4, and {iv} WS10 versus Late 4. In all scenarios, the MaxCombo test expectedly preserved the type I error rate of 5%. For the scenarios {i} – {iv} mentioned above, we found the empirical power of the MaxCombo test to be much higher than the target of 80% when $\mathrm{RT}(p_I) = 1.5$ and $\mathrm{RT}(p_{II}) = 2$. Contrastingly, we found the empirical power of the MaxCombo test to be much lower than the target of 80% when $\mathrm{RT}(p_I) = 2$ and $\mathrm{RT}(p_{II}) = 1.5$. In the latter case, the curves separate out early and this separation reduces over time (see Figure 2). However, all sample sizes in Table 2a (for $\beta_0 = 0.75, 1.0, 1.25$) fall in the "small to moderate" category (ranging from 20 to 100) and for moderately large sample sizes (ranging from 100 to 300, when $\beta_0 = 0.5$ in case of WS06 and WS10), the power of the MaxCombo did improve although it did not reach 80%.

To assess sensitivity to sample sizes, we conducted more simulations in Table 2b exclusively for $\beta_0 = 0.5$ with reduced effect size definitions (to achieve larger sample sizes ranging from 264 to 937) of $\mathrm{RT}(p_I) = 1.25, \mathrm{RT}(p_{II}) = 1.75$, and, $\mathrm{RT}(p_I) = 1.75, \mathrm{RT}(p_{II}) = 1.25$. In these simulations we considered the weighting schemes WS02, WS04, WS06, WS10 and WS12 as these result in different sample sizes using our WRT method. On doing so, we observed that the MaxCombo test showed very high power for $\mathrm{RT}(p_I) = 1.25, \mathrm{RT}(p_{II}) = 1.75$ when sample sizes were large (300+ resulting from WS02, WS04). The power went below 80% when sample sizes were relatively smaller (below 300 – resulting from WS10 and WS12). Analogously, when looking at $\mathrm{RT}(p_I) = 1.75, \mathrm{RT}(p_{II}) = 1.25$, the MaxCombo test did achieve adequate power in the case of large sample sizes (400+ resulting from WS10 and WS12), but very less power when sample sizes were relatively smaller (400- resulting from WS02 and WS04). All simulations for the MaxCombo method were conducted using the `NPH` package in the R statistical software[27].

Overall, our proposed WRT method maintained the type I error and achieved 80% power in all scenarios whereas the MaxCombo test always maintained type I error but needed large sample sizes to achieve 80% power. However, this should not be seen as a criticism of the MaxCombo method. It should be remembered that the MaxCombo test is considered a very robust alternative in the nonproportional hazards scenario if no preliminary information about the trial is available as noted in Lin et al.[28] and Roychoudhary et al.[29]. We chose to compare our WRT method to the MaxCombo for its popularity and recommended use by many authors – see also Dormuth et al.[30] for other omnibus test options. The adequacy of power for our WRT method is due to the preliminary knowledge we have about the Weibull shape parameters thereby allowing us to detect differences between two survival curves that follow specific monotonically increasing or decreasing treatment (longevity) effect.

### 4.3 *Sensitivity analysis using finer or coarser weighting schemes*

In addition to the 13 weighting schemes expressed through $u_j(p_j)$ for $j = 1,2, \dots ,9$ in Section 4.1, we conducted a sensitivity analysis using alternate weighting schemes. These included two alternate sets generated using – {a} $u_j(p_j)$ for $j = 1,2, \dots ,5$ and {b} $u_j(p_j)$ for $j = 1,2, \dots ,17$ and are as given below:

Set A: $u_j(p_j)$ for $j = 1,2, \dots ,5$ with $p_1 = 0.10, p_2 = 0.25, p_3 = 0.50, p_4 = 0.75, p_5 = 0.90$

| | |
|---|---|
| WS02a (Early effect important): | $u_j(p_j) \equiv 1, 1, 0, 0, 0$ |
| WS04a (Sharp decreasing effect): | $u_j(p_j) \equiv 9, 7.5, 5, 2.5, 1$ |
| WS06a (Equal weight to all percentiles): | $u_j(p_j) \equiv 1, 1, 1, 1, 1$ |
| WS10a (Sharp increasing effect): | $u_j(p_j) \equiv 1, 2.5, 5, 7.5, 9$ |
| WS12a (Late effect important): | $u_j(p_j) \equiv 0, 0, 0, 1, 1$ |

Set B: $u_j(p_j)$ for $j = 1,2, \dots ,17$ with $p_1 = 0.10,\ p_2 = 0.15,\ p_3 = 0.20,\ p_4 = 0.25,\ p_5 = 0.30,\ p_6 = 0.35,\ p_7 = 0.4,\ p_8 = 0.45,\ p_9 = 0.50,\ p_{10} = 0.55,\ p_{11} = 0.60,\ p_{12} = 0.65,\ p_{13} = 0.70,\ p_{14} = 0.75,\ p_{15} = 0.80,\ p_{16} = 0.85,\ p_{17} = 0.90$

| | |
|---|---|
| WS02b (Early effect important): | $u_j(p_j) \equiv 1,1,1,1,1,0,0,0,0,0,0,0,0,0,0,0,0$ |
| WS04b (Sharp decreasing effect): | $u_j(p_j) \equiv 9,8.5,8,7.5,7,6.5,6,5.5,5,4.5,4,3.5,3,2.5,2,1.5,1$ |
| WS06b (Equal weight to all p): | $u_j(p_j) \equiv 1,1,1,1,1,1,1,1,1,1,1,1,1,1,1,1,1$ |
| WS10b (Sharp increasing effect): | $u_j(p_j) \equiv 1,1.5,2,2.5,3,3.5,4,4.5,5,5.5,6,6.5,7,7.5,8,8.5,9$ |
| WS12b (Late effect important): | $u_j(p_j) \equiv 0,0,0,0,0,0,0,0,0,0,0,0,1,1,1,1,1$ |

Figure 4 displays the sample sizes for the four different scenarios represented in four panels by – {i} $\mathrm{RT}(p_I) = 1.25$ and $\mathrm{RT}(p_{II}) = 1.75$ {ii} $\mathrm{RT}(p_I) = 1.5$ and $\mathrm{RT}(p_{II}) = 2$ , {iii} $\mathrm{RT}(p_I) = 1.75$ and $\mathrm{RT}(p_{II}) = 1.25$, and {iv} $\mathrm{RT}(p_I) = 2$ and $\mathrm{RT}(p_{II}) = 1.5$. Four different values of $\beta_0 \equiv (0.5, 0.75, 1.0, 1.25)$ are considered in each panel and are represented by different colors. The other design specifications $\alpha = 0.05$, 80% (power), $t_{med_0} = 4$ months $a = 12$ months, $f = 12$ months, and $r = 1$. In all four panels and for all values of $\beta_0$, we see that the sample sizes are quite close for *Set B* (finer weighting with ventile increments between 0.1 and 0.9 – represented by dotted lines) compared to the *Reference Set* (weighting with decile increments between 0.1 and 0.9 – represented by solid lines). On the other hand, we see bigger differences in sample size between *Set A* (coarser weighting – represented by dashed lines) compared to the *Reference set* with most prominent differences arising when $\beta_0 < 1$ and for the weighting schemes that give more weight to early failures. Bigger sample size differences are also seen for smaller effect sizes of $\{i\}$ and $\{iii\}$ compared to $\{ii\}$ and $\{iv\}$. We can thus infer that there is no significant gain in precision of sample size calculation when using *Set B* compared to the *Reference set.* We thus propose to use the weighting schemes of the *Reference set* as the *default* option in our calculations – with the flexibility to change the *default* if needed.

## 5. Discussion

### 5.1 *Advantages of the WRT method*

We have proposed a new method called *Weighted Relative Time* to calculate sample size for two-arm trials with a time-to-event endpoint in the presence of NPH utilizing two different Weibull distributions. The biggest advantage of our proposed method is the *ease of interpretability* for both clinicians and patients who participate in the trial. This is because our method is based on the concept of *improvement in longevity* rather than *reduction in instantaneous risk* of an event – researchers can hypothesize an early and/or late benefit for the treatment under consideration versus control and can convey the same to potential participants in the trial. Once the data is collected and analyzed, the results can be easily conveyed through RT(p) for any percentile (say, $25^{th}$, median or $75^{th}$ percentile) along with the corresponding confidence interval. As shown in Section 4, the WRT method preserves the type I error and target power even in the case of small samples in addition to having other satisfactory operation characteristics. It allows incorporation of very flexible weighting schemes that allot increased focus to early, middle or late failures. While the weighted logrank test and the MaxCombo test incorporate weights for the difference between observed and expected events at event times, our WRT method contrastingly assigns weights to the percentiles of survival times. The previously published method of Phadnis and Mayo[10], a test of medians, PH sample size formula of Schoenfeld, and testing for two groups using the exponential distribution are all special cases of the WRT method. The special case of *equal weights* is also connected to the RMST approach. Importantly, our method encourages more communication between researchers and statisticians at the design phase of an intended trial. Statisticians can elicit information from their research collaborators by asking questions about the differential effect of a treatment rather than assume a single constant effect (such as *reduction in HR* or *improvement in median*). Although research collaborators may have some idea about how treatment benefit changes over time, they may not mention this to a statistician unless they ask for it.

5.2 *Limitations*

The major limitation of the WRT method is that the control arm shape parameter $\beta_0$ and treatment arm shape parameter $\beta_1$ are assumed fixed and the theory of the test statistic $Q_w$ requires this assumption. When $\beta_0$ and $\beta_1$ are not fixed, they will have to be estimated from data and the additional uncertainty around this estimation will require inflation of the sample size proposed by the WRT method. One solution would be to use this sample size as an initial estimate and study the operation characteristics via simulations. If the operation characteristics are found unsatisfactory, the sample size can be increased, and the operation characteristics can be assessed again. Future work in this direction can be conducted in a Bayesian setting.

Another limitation is that it is restricted to NPH resulting from two different Weibull distributions and more work needs to be carried out to extend it to other parametric distributions. The WRT method does not allow crossing of curves except at $q_{min}$ (see Figure 1) and $q_{max}$ (see Figure 2). Additionally, it is developed for monotonically increasing or monotonically decreasing time ratios at the survival percentiles $p_j$ in the region enclosed by $p_I$ and $p_{II}$. Further, uniform accrual is assumed in the sample size calculations.

5.3 *Practical considerations*

Researchers and statisticians are expected to be aware of certain practical considerations when designing a trial using the WRT method. First, the choice of the weighting scheme should be chosen when formulating research hypotheses and this should be mentioned in the statistical section of the research protocol. If not done so, there could be a tendency for *data snooping* where in, after the trial data is available, a statistician (or his collaborator) may be tempted to use a weighting scheme that is more likely to reject the null hypotheses. Standard statistical texts warn of a similar situation in the context of using the weighted logrank test where the choice of FH weights are recommended to be chosen well in advance of the trial data analysis. Second, researchers need to come up with a practically relevant (driven by clinical considerations) choice of $p_I$ and $p_{II}$ when defining the effect sizes $RT(p_I)$ and $RT(p_{II})$ respectively. As an example, say $p_I = 0.1$ and $p_{II} = 0.9$ are chosen. But the study accrual time $a$ and follow-up time $f$ is such that $1 - S(a + f) = 0.8$, then $RT(p_{II})$ will never be realized in the actual trial. Thus, if the researchers suspect that total study time $t_{total} = a + f$ is relatively short, they may have to choose longer accrual and/or follow-up times to ensure that $p_{II} < p_{max} = 1 - S(a + f)$. Additionally, in scenarios where $RT(p_I) < RT(p_{II})$ is true, researchers will have to ensure that their choice of $p_I > q_{min}$ is such that the value of $q_{min}$ is acceptable from a clinical viewpoint. Analogously, in scenarios where $RT(p_I) > RT(p_{II})$ is true, researchers will have to ensure that their choice of $p_{II} < q_{max}$ is such that the value of $q_{max}$ represents a clinically relevant threshold.

Overall, the WRT method is an important addition to a statistician's arsenal for designing trials in the case of nonproportional hazards with good prospects for extending it to more advanced fixed and adaptive (frequentist or Bayesian) settings.

**Acknowledgements:** The high-performance computing capabilities used to conduct some of the simulations described in this paper, were supported in part by the National Cancer Institute (NCI) Cancer Center Support Grant P30 CA168524.

## Appendix A Theory and framework

### A.1 Concept of Relative Time RT(p)

The Weibull distribution is a two-parameter distribution whose probability density function is:

$$f(t) = \frac{\beta}{\theta^{\beta}} t^{\beta-1} \exp\{-(t/\theta)^{\beta}\} \qquad \theta, \beta > 0, t > 0 \tag{A.1.1}$$

Here, $\theta$ is a scale parameter and $\beta$ is a shape parameter that determines the shape of the hazard function ($\beta > 1$ gives hazard that increases over time, $\beta < 1$ gives hazard that decreases over time, and $\beta = 1$ representing the special case of exponential distribution with constant hazard). Relative time can be defined as the ratio of times at which exactly *p* % of the individuals in one study arm experience an event of interest. Due to the dependence on *p*, it is denoted as RT(p). Thus, the interpretation of RT(p) is that the time required for *p* % of the individuals in one study arm to experience an event is RT(p)-fold times the time required for *p* % of the individuals in other study arm. That is,

$$RT(p) = \frac{t_1(p)}{t_0(p)} = \frac{S_1^{-1}(1-p)}{S_0^{-1}(1-p)} \tag{A.1.2}$$

Here, $S_i^{-1}(1-p)$ is the inverse survival function for study arm *i* (*i* = 0, 1). Let $\theta_i$ and $\beta_i$ (for *i* = 0, 1) represent the scale and shape parameters of two different Weibull distributions for control arm C and new experimental treatment arm E. From (A.1.2) we also have $1 - p = S_i\{t_i(p)\}$ and from (1) we know that $S_i\{t_i(p)\} = \exp\left[-\{t_i(p)/\theta_i\}^{\beta_i}\right]$ leading to

$$t_i(p) = \theta_i \left\{\log\left(\frac{1}{1-p}\right)\right\}^{1/\beta_i} \tag{A.1.3}$$

Thus RT(p) can be expressed as:

$$RT(p) = \frac{t_1(p)}{t_0(p)} = \frac{\theta_1}{\theta_0}\left\{\log\left(\frac{1}{1-p}\right)\right\}^{\frac{1}{\beta_1}-\frac{1}{\beta_0}} \tag{A.1.4}$$

On the logarithm scale, equation (4) becomes

$$\log\{RT(p)\} = \{\log(\theta_1) - \log(\theta_0)\} + \left(\frac{1}{\beta_1} - \frac{1}{\beta_0}\right). \log\left\{\log\left(\frac{1}{1-p}\right)\right\} \qquad (\mathrm{A}.1.5)$$

From equations (A.1.4) and (A.1.5) we see that when $\beta_1 = \beta_0$, the dependency on $p$ disappears and in that case RT(p) is a constant and can be called 'Proportional Time (PT)'. In case of Weibull, this also results in the PH assumption being true, but this is not true for other distributions. See Cox et al. (2007) for more details on this topic where the Weibull is shown as a special case of the generalized gamma distribution.

Before moving to the next section on test statistic development, it is important to discuss the notion of effect size as regards to the concept of Relative Time $RT(p)$ discussed above. When designing trials with the usual PH assumption or the assumption of exponentially distributed survival times, the effect size is often defined as the improvement in median survival times by a factor $c$. In our context, this amounts to the following information needed to define the effect size: $p = 0.5$ ($50^{\text{th}}$ percentile) and $RT(p) = c$. However, in our case we are concerned about non-PH and non-PT and this can be interpreted as *"hazard ratio changes over time"* or *"time-ratio (acceleration factor) changes over time"*. That is, to define a '*non-constant effect of treatment',* we require knowledge of two coordinates – $\{p_I, RT(p_I)\}$ and $\{p_{II}, RT(p_{II})\}$. The concept of Relative Time can now be extended in the following way:

For $i = 1,2$, the time required for $p_i$ % of the individuals in one study arm to experience an event is $RT(p_i)$-fold times the time required for $p_i$ % of the individuals in other study arm. Thus $p_I$ is an "early" quantile at which researchers anticipate an improvement in life course for new treatment versus control denoted by $RT(p_I)$, and $p_{II}$ is an "later" quantile at which researchers anticipate an improvement in life course for new treatment versus control denoted by $RT(p_{II})$. The examples in Section 2 have $p_I = 0.10$ ($10^{\text{th}}$ percentile) and $p_{II} = 0.90$ ($90^{\text{th}}$ percentile.

## A.2 Test statistic development

Consider the hypotheses of Phadnis and Mayo[10] given by:

$$\mathrm{H}_0:\ RT(p_{mid}) \le 1$$

$$\mathrm{H}_1:\ RT(p_{mid}) > 1$$

Here, $p_{mid}$ can be any user-defined value of $p$ that falls in the *middle* of $p_I$ and $p_{II}$. By default (if not user-defined), it can be taken as $p_{mid} = (p_I + p_{II})/2$ , the average of the two quantiles at which the effect size is defined. Let $\hat{\theta}_i$ for $i$ = 0, 1 be the maximum likelihood estimate of $\theta_i$ in the C and E arms respectively. Then we know that for the Weibull distribution with all observations as events (no censoring) with $d_i$ events in the $i^{\text{th}}$ arm, we have $\hat{\theta}_i = \left(\frac{\sum_{j=1}^{d_i} t_{ij}^{\beta_i}}{d_i}\right)^{1/\beta_i}$.

Since each $t_{ij} \sim \text{Weibull}(\theta_i, \beta_i)$, it can be shown that $\hat{\theta}_i \sim \text{GG}\left(d_i\,, \theta_i/{d_i}^{1/\beta_i}\,, \beta_i\right)$ where the letters GG stand for a 3-parameter generalized gamma distribution. From this we get

$$\hat{\theta}_i\left\{\log\left(\frac{1}{1-p_{mid}}\right)\right\}^{\frac{1}{\beta_i}} \sim \text{GG}\left[d_i\,, \theta_i\left\{\frac{\log\left(\frac{1}{1-p_{mid}}\right)}{d_i}\right\}^{\frac{1}{\beta_i}}, \beta_i\right] \qquad (\text{A}.2.1)$$

Using a reparameterization by taking $\lambda_i = 1/\sqrt{d_i}$, $\sigma_i = 1/\left(\beta_i\sqrt{d_i}\right)$ and

$\mu_i = \log\left[\theta_i\left\{\log\left(\frac{1}{1-p_{mid}}\right)\right\}^{\frac{1}{\beta_i}}\right]$ we can say that $\hat{\theta}_i[\log\{1/(1-p_{mid})\}]^{1/\beta_i} \sim \text{GG}(\lambda_i\,, \mu_i, \sigma_i)$. The advantage of this reparameterization is that we can see that as the number of events $d_i$ increases, $\lambda_i$ decreases towards 0. Even for $d_i = 25$, we get $\lambda_i = 0.2$. A well-known property of the GG distribution is that as $\lambda_i \to 0$, the distribution converges to a lognormal distribution. That is, denoting $Q_i = \hat{\theta}_i[\log\{1/(1-p_{mid})\}]^{1/\beta_i}$ , we get $Q_i \sim \text{Lognormal}(\mu_i, \sigma_i)$. See Stacy and Mihram (1965) for more properties of the GG distribution and Cox et al. (2007) explaining how some popular distributions such as Weibull, lognormal, gamma, ammag, inverse Weibull, inverse gamma, and exponential are special cases of the GG distribution.

Following the above, Phadnis and Mayo[10] have developed a test statistic that asymptotically follows a normal distribution $Q' = \log(Q_1/Q_0) \sim \mathrm{N}(\mu_d, \sigma_d^2)$ with $\mu_d = \log\left[\frac{\theta_1}{\theta_0}\left\{\log\left(\frac{1}{1-p_{mid}}\right)\right\}^{\frac{1}{\beta_1}-\frac{1}{\beta_0}}\right]$ and $\sigma_d = \sqrt{\frac{1}{d_1\beta_1^2}+\frac{1}{d_0\beta_0^2}}$ allowing them to calculate the number of events in the control arm $d_0$ and treatment arm $d_1$ using the allocation ratio $r$ for a one-sided test with the formula

$$d_0 = \left[\frac{(Z_\omega + Z_{1-\alpha})}{\log\{RT(p_{mid})\}}\right]^2 \left(\frac{1}{r\beta_1^2}+\frac{1}{\beta_0^2}\right)$$

$$d_1 = rd_0 \tag{A.2.2}$$

where $\alpha$ is the type I error for the one-sided test and $1-\omega$ is the target power of the test. For a two-sided test, we can use $\alpha/2$ in place of $\alpha$. Phadnis and Mayo[10] further discuss the calculation of sample sizes in the C and E arms by first adjusting for administrative censoring, and second, adjusting for loss to follow-up by reformulating using an Extended Cox model. Their simulations show that in most scenarios, the empirical type I error and power at maintained at the desired levels.

### A.3 Calculation of sample size accounting for administrative censoring

Assuming a uniform accrual, the censoring distribution function $G(t)$ is given by

$$G(t) = \begin{cases} 1 & \text{if } t \le f \\ \dfrac{a+f-t}{a} & \text{if } f \le t \le a+f \\ 0 & \text{otherwise} \end{cases} \tag{A.3.1}$$

where $a$ and $f$ are the accrual and follow-up time respectively. Then the probability that a subject experiences an event during the trial in arm $i$ ($i$ = 0, 1) is

$$v_i = \int_0^{\infty} G(t).f_i(t)dt \qquad (A.3.2)$$

where $f_i(t)$ is $f(t)$ with $\theta = \theta_i$ and $\beta = \beta_i$. Dividing the number of events obtained through the sample size formula in (A.2.2) by $v = (v_0 + v_1)/2$ gives the sample size adjusted for administrative censoring. Alternatively, $v_i$ can be calculated using Simpson's rule by

$$v_i = 1 - \frac{1}{6}\{S_i(f) + 4S_i(f + 0.5a) + S_i(f + a)\} \qquad (A.3.3)$$

where $S_i(t)$ is the survival function of the Weibull with $\theta = \theta_i$ and $\beta = \beta_i$.

### A.4 Correlation $\hat{\rho}_{ab}$ of Log-Relative Time values from a sample

The estimated log-relative time:

$$\log \widehat{RT}(p) = \log \hat{\theta}_t - \log \hat{\theta}_c + \left(\frac{1}{\beta_t} - \frac{1}{\beta_c}\right) \log\{-\log(1 - p)\}$$

Here, $\beta_t$ and $\beta_c$ are fixed. Define $A = log\ \hat{\theta}_t - log\ \hat{\theta}_c$ and $k(p) = (\frac{1}{\beta_t} - \frac{1}{\beta_c})\ log\{-log(1 - p)\}$ so that, $\log \widehat{RT}(p) = A + k(p)$. Here $A$ captures all stochastic variation and $k(p)$ is a deterministic function of $p$. For any two quantiles $p_a, p_b \in (0,1)$, their estimates differ by the fixed constant $k(p_b) - k(p_a)$.

$$log\ \widehat{RT}(p_b) = log\ \widehat{RT}(p_a) + \{k(p_b) - k(p_a)\}.$$

Hence, by invariance of the Pearson correlation under additive location shifts,

$\hat{\rho}_{ab} = Corr[log\ \widehat{RT}(p_a), log\ \widehat{RT}(p_b)] = Corr(A + k(p_a), A + k(p_b)) = Corr(A, A) = 1.$

Table 1. Event size/Sample size in each arm (given by d/n) for proposed method with $\alpha = 0.05$, one-sided test, 80% power for $a = f = 12$, allocation ratio $r = 1$ for varying combinations of control-arm shape parameter $\beta_0$ with $t_{median,0} = 4$, and four effect size definitions specified by RT($p_I$) and RT($p_{II}$) with $p_I = 0.10,\ p_{II} = 0.90$ using different Weighting Schemes

| Shape of control arm | Weighing Schemes | $RT(p_I) = 1.5; RT(p_{II}) = 2.0$ | | $RT(p_I) = 1.25; RT(p_{II}) = 1.75$ | | $RT(p_I) = 2; RT(p_{II}) = 1.5$ | | $RT(p_I) = 1.75; RT(p_{II}) = 1.25$ | |
|---|---|---|---|---|---|---|---|---|---|
| | | $\beta_1$ | d/n | $\beta_1$ | d/n | $\beta_1$ | d/n | $\beta_1$ | d/n |
| | WS 01 | | 316/ 444 | | 1049/ 1450 | | 99/ 136 | | 150/ 202 |
| | WS 02 | | 238/ 335 | | 602/ 831 | | 119/ 164 | | 197/ 266 |
| | WS 03 | | 190/ 266 | | 401/ 554 | | 143/ 197 | | 264/ 356 |
| | WS 04 | | 188/ 264 | | 396/ 547 | | 144/ 199 | | 267/ 360 |
| | WS 05 | | 180/ 253 | | 368/ 508 | | 150/ 207 | | 285/ 385 |
| | WS 06 | | 160/ 225 | | 304/ 419 | 0.5245 | 169/ 233 | 0.5289 | 347/ 468 |
| $\beta_0 = 0.50$ | WS 07 | 0.4777 | 154/ 217 | 0.4741 | 286/ 395 | | 176/ 243 | | 372/ 502 |
| | WS 08 | | 154/ 216 | | 285/ 393 | | 177/ 244 | | 374/ 505 |
| | WS 09 | | 143/ 201 | | 255/ 352 | | 192/ 265 | | 432/ 582 |
| | WS 10 | | 138/ 194 | | 240/ 332 | | 201/ 278 | | 470/ 633 |
| | WS 11 | | 135/ 189 | | 232/ 320 | | 208/ 286 | | 496/ 669 |
| | WS 12 | | 119/ 167 | | 191/ 264 | | 248/ 342 | | 695/ 937 |
| | WS 13 | | 108/ 152 | | 167/ 231 | | 288/ 397 | | 941/ 1270 |
| | WS 01 | | 144/ 179 | | 480/ 585 | | 43/ 52 | | 65/ 77 |
| | WS 02 | | 109/ 135 | | 275/ 336 | | 52/ 63 | | 86/ 101 |
| | WS 03 | | 87/ 108 | | 183/ 224 | | 62/ 75 | | 115/ 135 |
| | WS 04 | | 86/ 107 | | 181/ 221 | | 63/ 76 | | 116/ 136 |
| | WS 05 | | 82/ 102 | | 168/ 205 | | 66/ 79 | | 124/ 145 |
| | WS 06 | | 73/ 91 | | 139/ 169 | | 74/ 89 | | 150/ 177 |
| $\beta_0 = 0.75$ | WS 07 | 0.7010 | 70/ 88 | 0.6933 | 131/ 160 | 0.8064 | 77/ 93 | 0.8168 | 161/ 190 |
| | WS 08 | | 70/ 88 | | 130/ 159 | | 77/ 93 | | 162/ 191 |
| | WS 09 | | 66/ 82 | | 117/ 142 | | 84/ 101 | | 187/ 220 |
| | WS 10 | | 63/ 79 | | 110/ 134 | | 88/ 106 | | 203/ 239 |
| | WS 11 | | 62/ 77 | | 106/ 129 | | 90/ 109 | | 215/ 253 |
| | WS 12 | | 54/ 68 | | 88/ 107 | | 108/ 130 | | 301/ 354 |
| | WS 13 | | 50/ 62 | | 77/ 94 | | 125/ 151 | | 407/ 480 |
| | WS 01 | | 83/ 95 | | 277/ 312 | | 24/ 26 | | 36/ 39 |
| | WS 02 | | 63/ 72 | | 159/ 179 | | 29/ 32 | | 47/ 51 |
| | WS 03 | | 50/ 57 | | 106/ 119 | | 35/ 38 | | 63/ 68 |
| | WS 04 | | 50/ 57 | | 105/ 118 | | 35/ 38 | | 64/ 69 |

| | | | | | | | | |
|---|---|---|---|---|---|---|---|---|
| | WS 05 | | 48/ 55 | | 97/ 110 | | 36/ 40 | | 68/ 73 |
| | WS 06 | | 42/ 49 | | 80/ 90 | | 41/ 45 | | 83/ 89 |
| $\beta_0 = 1.00$ | WS 07 | 0.9147 | 41/ 47 | 0.9016 | 76/ 85 | 1.1029 | 42/ 47 | 1.1225 | 88/ 95 |
| | WS 08 | | 41/ 47 | | 76/ 85 | | 43/ 47 | | 89/ 96 |
| | WS 09 | | 38/ 44 | | 68/ 76 | | 46/ 51 | | 103/ 111 |
| | WS 10 | | 37/ 42 | | 64/ 72 | | 48/ 54 | | 112/ 120 |
| | WS 11 | | 36/ 41 | | 62/ 69 | | 50/ 55 | | 118/ 127 |
| | WS 12 | | 31/ 36 | | 51/ 57 | | 60/ 66 | | 165/ 178 |
| | WS 13 | | 29/ 33 | | 45/ 50 | | 69/ 76 | | 223/ 241 |
| | WS 01 | | 55/ 60 | | 183/ 196 | | 15/ 16 | | 23/ 23 |
| | WS 02 | | 41/ 45 | | 105/ 112 | | 18/ 19 | | 30/ 30 |
| | WS 03 | | 33/ 36 | | 70/ 75 | | 22/ 23 | | 39/ 41 |
| | WS 04 | | 33/ 36 | | 69/ 74 | | 22/ 23 | | 40/ 41 |
| | WS 05 | | 31/ 34 | | 64/ 69 | | 23/ 24 | | 42/ 44 |
| | WS 06 | | 28/ 30 | | 53/ 57 | | 26/ 27 | | 52/ 53 |
| $\beta_0 = 1.25$ | WS 07 | 1.1195 | 27/ 29 | 1.1000 | 50/ 54 | 1.4150 | 27/ 28 | 1.4474 | 55/ 57 |
| | WS 08 | | 27/ 29 | | 50/ 53 | | 27/ 28 | | 56/ 57 |
| | WS 09 | | 25/ 27 | | 45/ 48 | | 29/ 31 | | 64/ 66 |
| | WS 10 | | 24/ 26 | | 42/ 45 | | 30/ 32 | | 70/ 72 |
| | WS 11 | | 24/ 26 | | 41/ 43 | | 31/ 33 | | 74/ 76 |
| | WS 12 | | 21/ 23 | | 34/ 36 | | 37/ 39 | | 103/ 106 |
| | WS 13 | | 19/ 21 | | 29/ 32 | | 43/ 45 | | 139/ 143 |

Table 2a. Empirical type I error and power for different settings of the proposed method for one-sided test with $a$ = 12, f = 12, $p_I = 0.10,\ p_{II} = 0.90$) from 10,000 simulations for 1:1 randomization in the two study arms versus empirical type I error and power for different settings of the Max Combo test (Broad 6, Classic 4, Early 4 and Late 4)

| Proposed Method (Weighted Relative Time) | | | | | | | | | | Max Combo Test | | | | | | | |
|---|---|---|---|---|---|---|---|---|---|---|---|---|---|---|---|---|---|
| Weight Scheme | Control Shape $\beta_0$ | Treatment Shape $\beta_1$ | Effect Size | Events/ S size d/n | Emp. Type I | Emp. Power | ARB % | MSE | Coverage % | Broad 6 Emp. Type I | Broad 6 Emp. Power | Classic 4 Emp. Type I | Classic 4 Emp. Power | Early 4 Emp. Type I | Early 4 Emp. Power | Late 4 Emp. Type I | Late 4 Emp. Power |
| WS04 Early effects are more important | 0.50 | 0.4777 | $RT(p_I) = 1.5$<br>$RT(p_{II}) = 2.0$ | 188/264 | 0.039 | 0.799 | -0.841 | 0.0443 | 90.36 | 0.048 | 0.869 | 0.048 | 0.883 | 0.049 | 0.863 | 0.049 | 0.880 |
| | | 0.5245 | $RT(p_I) = 2.0$<br>$RT(p_{II}) = 1.5$ | 144/199 | 0.049 | 0.797 | -0.194 | 0.0538 | 90.04 | 0.051 | 0.648 | 0.050 | 0.662 | 0.051 | 0.691 | 0.049 | 0.624 |
| | 0.75 | 0.7010 | $RT(p_I) = 1.5$<br>$RT(p_{II}) = 2.0$ | 86/107 | 0.038 | 0.803 | 0.298 | 0.0454 | 89.91 | 0.050 | 0.884 | 0.050 | 0.898 | 0.050 | 0.876 | 0.052 | 0.895 |
| | | 0.8064 | $RT(p_I) = 2.0$<br>$RT(p_{II}) = 1.5$ | 63/76 | 0.050 | 0.800 | -0.584 | 0.0528 | 89.99 | 0.050 | 0.621 | 0.049 | 0.636 | 0.050 | 0.663 | 0.053 | 0.595 |
| | 1.00 | 0.9147 | $RT(p_I) = 1.5$<br>$RT(p_{II}) = 2.0$ | 50/57 | 0.038 | 0.808 | 0.398 | 0.0454 | 89.71 | 0.054 | 0.892 | 0.053 | 0.905 | 0.051 | 0.881 | 0.054 | 0.909 |
| | | 1.1029 | $RT(p_I) = 2.0$<br>$RT(p_{II}) = 1.5$ | 35/38 | 0.054 | 0.801 | -0.364 | 0.0542 | 89.93 | 0.057 | 0.595 | 0.051 | 0.613 | 0.053 | 0.635 | 0.057 | 0.571 |
| | 1.25 | 1.1195 | $RT(p_I) = 1.5$<br>$RT(p_{II}) = 2.0$ | 33/36 | 0.035 | 0.806 | 0.113 | 0.0450 | 89.82 | 0.055 | 0.897 | 0.052 | 0.907 | 0.052 | 0.879 | 0.057 | 0.911 |
| | | 1.4150 | $RT(p_I) = 2.0$<br>$RT(p_{II}) = 1.5$ | 22/23 | 0.056 | 0.803 | -0.397 | 0.0528 | 90.25 | 0.061 | 0.586 | 0.057 | 0.598 | 0.053 | 0.620 | 0.062 | 0.565 |
| WS06 Equal weights | 0.50 | 0.4777 | $RT(p_I) = 1.5$<br>$RT(p_{II}) = 2.0$ | 160/225 | 0.041 | 0.802 | -0.382 | 0.0526 | 90.09 | 0.047 | 0.818 | 0.047 | 0.834 | 0.050 | 0.810 | 0.047 | 0.832 |
| | | 0.5245 | $RT(p_I) = 2.0$<br>$RT(p_{II}) = 1.5$ | 169/233 | 0.047 | 0.800 | -0.602 | 0.0458 | 90.26 | 0.050 | 0.706 | 0.049 | 0.717 | 0.049 | 0.745 | 0.050 | 0.683 |
| | 0.75 | 0.7010 | $RT(p_I) = 1.5$<br>$RT(p_{II}) = 2.0$ | 73/91 | 0.048 | 0.803 | 0.247 | 0.0543 | 89.00 | 0.054 | 0.830 | 0.054 | 0.848 | 0.055 | 0.821 | 0.054 | 0.846 |
| | | 0.8064 | $RT(p_I) = 2.0$<br>$RT(p_{II}) = 1.5$ | 74/89 | 0.050 | 0.799 | -0.635 | 0.0458 | 90.01 | 0.050 | 0.677 | 0.051 | 0.695 | 0.050 | 0.717 | 0.051 | 0.651 |
| | 1.00 | 0.9147 | $RT(p_I) = 1.5$<br>$RT(p_{II}) = 2.0$ | 42/49 | 0.036 | 0.801 | -0.116 | 0.0530 | 89.95 | 0.053 | 0.844 | 0.052 | 0.858 | 0.050 | 0.829 | 0.054 | 0.863 |
| | | 1.1029 | $RT(p_I) = 2.0$<br>$RT(p_{II}) = 1.5$ | 41/45 | 0.054 | 0.804 | -0.081 | 0.0448 | 90.09 | 0.053 | 0.662 | 0.053 | 0.677 | 0.050 | 0.706 | 0.055 | 0.638 |
| | 1.25 | 1.1195 | $RT(p_I) = 1.5$<br>$RT(p_{II}) = 2.0$ | 28/30 | 0.037 | 0.795 | -0.247 | 0.0536 | 89.63 | 0.059 | 0.840 | 0.057 | 0.854 | 0.058 | 0.817 | 0.061 | 0.862 |
| | | 1.4150 | $RT(p_I) = 2.0$<br>$RT(p_{II}) = 1.5$ | 26/27 | 0.059 | 0.807 | -0.333 | 0.0449 | 89.78 | 0.059 | 0.653 | 0.056 | 0.669 | 0.053 | 0.695 | 0.061 | 0.633 |
| WS10 Late Effects are more important | 0.50 | 0.4777 | $\mathrm{RT}(p_I) = 1.5$<br>$\mathrm{RT}(p_{II}) = 2.0$ | 138/194 | 0.041 | 0.802 | -0.020 | 0.0617 | 89.63 | 0.050 | 0.764 | 0.052 | 0.785 | 0.053 | 0.755 | 0.052 | 0.783 |
| | | 0.5245 | $\mathrm{RT}(p_I) = 2.0$<br>$\mathrm{RT}(p_{II}) = 1.5$ | 201/278 | 0.048 | 0.804 | -0.131 | 0.0377 | 90.17 | 0.051 | 0.779 | 0.051 | 0.793 | 0.051 | 0.817 | 0.051 | 0.760 |
| | 0.75 | 0.7010 | $\mathrm{RT}(p_I) = 1.5$<br>$\mathrm{RT}(p_{II}) = 2.0$ | 63/79 | 0.037 | 0.800 | 0.451 | 0.0625 | 89.41 | 0.049 | 0.783 | 0.051 | 0.800 | 0.052 | 0.769 | 0.052 | 0.799 |
| | | 0.8064 | $\mathrm{RT}(p_I) = 2.0$<br>$\mathrm{RT}(p_{II}) = 1.5$ | 88/106 | 0.052 | 0.802 | -0.047 | 0.0381 | 89.85 | 0.054 | 0.744 | 0.054 | 0.760 | 0.053 | 0.786 | 0.055 | 0.722 |
| | 1.00 | 0.9147 | $\mathrm{RT}(p_I) = 1.5$<br>$\mathrm{RT}(p_{II}) = 2.0$ | 37/42 | 0.036 | 0.801 | 0.408 | 0.0620 | 89.69 | 0.055 | 0.792 | 0.055 | 0.809 | 0.051 | 0.775 | 0.058 | 0.812 |
| | | 1.1029 | $\mathrm{RT}(p_I) = 2.0$<br>$\mathrm{RT}(p_{II}) = 1.5$ | 48/54 | 0.056 | 0.801 | 0.578 | 0.0376 | 89.98 | 0.056 | 0.740 | 0.056 | 0.753 | 0.054 | 0.779 | 0.056 | 0.716 |
| | 1.25 | 1.1195 | $\mathrm{RT}(p_I) = 1.5$<br>$\mathrm{RT}(p_{II}) = 2.0$ | 24/26 | 0.038 | 0.809 | 1.184 | 0.0627 | 89.59 | 0.060 | 0.800 | 0.056 | 0.816 | 0.054 | 0.773 | 0.062 | 0.823 |
| | | 1.4150 | $\mathrm{RT}(p_I) = 2.0$<br>$\mathrm{RT}(p_{II}) = 1.5$ | 30/32 | 0.055 | 0.801 | 0.604 | 0.0370 | 90.42 | 0.057 | 0.723 | 0.053 | 0.734 | 0.051 | 0.758 | 0.058 | 0.700 |

* The Broad 6 set includes the components FH(0,0), FH(1,0), FH(0,1), FH(1,1), FH(2,0) and FH(0,2). The Classic 4 set includes the components FH(0,0), FH(1,0), FH(0,1) and FH(1,1). The Early 4 set includes the components FH(0,0), FH(0.5, 0), FH(1,0) and FH(2,0). The Late 4 set includes the components FH(0,0), FH(0, 0.5), FH(0,1) and FH(0,2).

Table 2b. *Large Sample Size Scenario:* Empirical type I error and power for proposed method for one-sided test when $\beta_0 = 0.5$ with $a = 12$, f = 12, $p_I = 0.10$, $p_{II} = 0.90$) from 10,000 simulations for 1:1 randomization versus empirical type I error and power for different settings of the Max Combo test (Broad 6, Classic 4, Early 4 and Late 4)

| Proposed Method (Weighted Relative Time) | | | | | | | | | | Max Combo Test | | | | | | | |
|---|---|---|---|---|---|---|---|---|---|---|---|---|---|---|---|---|---|
| Control Shape $\beta_0$ | Weight Scheme | Treatment Shape $\beta_1$ | Effect Size | Events/ S size d/n | Emp. Type I | Emp. Power | ARB % | MSE | Coverage % | Broad 6 Emp. Type I | Broad 6 Emp. Power | Classic 4 Emp. Type I | Classic 4 Emp. Power | Early 4 Emp. Type I | Early 4 Emp. Power | Late 4 Emp. Type I | Late 4 Emp. Power |
| 0.5 | WS02 | 0.4741 | $RT(p_I) = 1.25$ $RT(p_{II}) = 1.75$ | 602/831 | 0.041 | 0.798 | -0.569 | 0.0143 | 89.85 | 0.048 | 0.985 | 0.049 | 0.988 | 0.050 | 0.980 | 0.049 | 0.988 |
| | | 0.5289 | $RT(p_I) = 1.75$ $RT(p_{II}) = 1.25$ | 197/266 | 0.049 | 0.801 | -0.546 | 0.0381 | 90.33 | 0.048 | 0.465 | 0.049 | 0.469 | 0.052 | 0.518 | 0.048 | 0.411 |
| | WS04 | 0.4741 | $RT(p_I) = 1.25$ $RT(p_{II}) = 1.75$ | 396/547 | 0.041 | 0.790 | -1.800 | 0.0217 | 89.93 | 0.051 | 0.925 | 0.050 | 0.936 | 0.050 | 0.909 | 0.049 | 0.936 |
| | | 0.5289 | $RT(p_I) = 1.75$ $RT(p_{II}) = 1.25$ | 267/360 | 0.049 | 0.800 | -1.112 | 0.0284 | 90.45 | 0.048 | 0.580 | 0.048 | 0.585 | 0.047 | 0.638 | 0.048 | 0.523 |
| | WS06 | 0.4741 | $RT(p_I) = 1.25$ $RT(p_{II}) = 1.75$ | 304/419 | 0.042 | 0.803 | -0.773 | 0.0282 | 89.78 | 0.052 | 0.851 | 0.052 | 0.868 | 0.051 | 0.836 | 0.049 | 0.867 |
| | | 0.5289 | $RT(p_I) = 1.75$ $RT(p_{II}) = 1.25$ | 347/468 | 0.053 | 0.807 | -0.565 | 0.0217 | 90.18 | 0.052 | 0.684 | 0.051 | 0.690 | 0.049 | 0.737 | 0.053 | 0.629 |
| | WS10 | 0.4741 | $RT(p_I) = 1.25$ $RT(p_{II}) = 1.75$ | 240/332 | 0.042 | 0.798 | -0.759 | 0.0357 | 90.05 | 0.052 | 0.770 | 0.052 | 0.791 | 0.052 | 0.743 | 0.052 | 0.792 |
| | | 0.5289 | $RT(p_I) = 1.75$ $RT(p_{II}) = 1.25$ | 470/633 | 0.054 | 0.802 | -0.640 | 0.0161 | 89.86 | 0.051 | 0.801 | 0.051 | 0.803 | 0.052 | 0.843 | 0.052 | 0.742 |
| | WS12 | 0.4741 | $RT(p_I) = 1.25$ $RT(p_{II}) = 1.75$ | 191/264 | 0.040 | 0.800 | -0.346 | 0.0438 | 90.17 | 0.048 | 0.682 | 0.049 | 0.704 | 0.049 | 0.654 | 0.049 | 0.701 |
| | | 0.5289 | $RT(p_I) = 1.75$ $RT(p_{II}) = 1.25$ | 695/937 | 0.048 | 0.802 | 0.654 | 0.0110 | 89.96 | 0.048 | 0.914 | 0.049 | 0.916 | 0.050 | 0.939 | 0.050 | 0.878 |

* The Broad 6 set includes the components FH(0,0), FH(1,0), FH(0,1), FH(1,1), FH(2,0) and FH(0,2). The Classic 4 set includes the components FH(0,0), FH(1,0), FH(0,1) and FH(1,1). The Early 4 set includes the components FH(0,0), FH(0.5, 0), FH(1,0) and FH(2,0). The Late 4 set includes the components FH(0,0), FH(0, 0.5), FH(0,1) and FH(0,2).

List of Figures:

Figure 1. Monotonically increasing treatment effect with RT(0.1) = 1.5 and RT(0.9) = 2.0

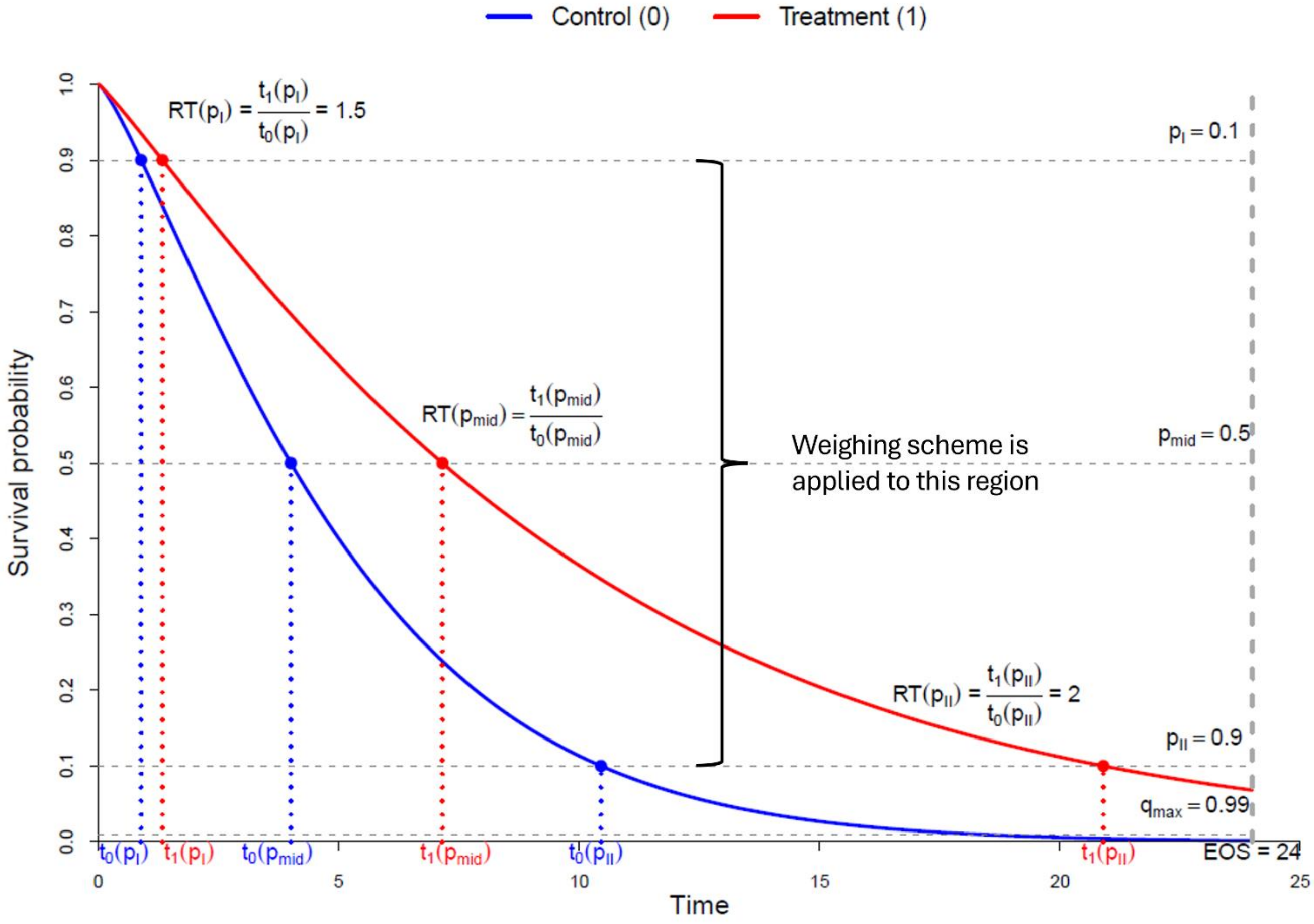

Figure 2. Monotonically decreasing treatment effect with RT(0.1) = 2.0 and RT(0.9) = 1.5

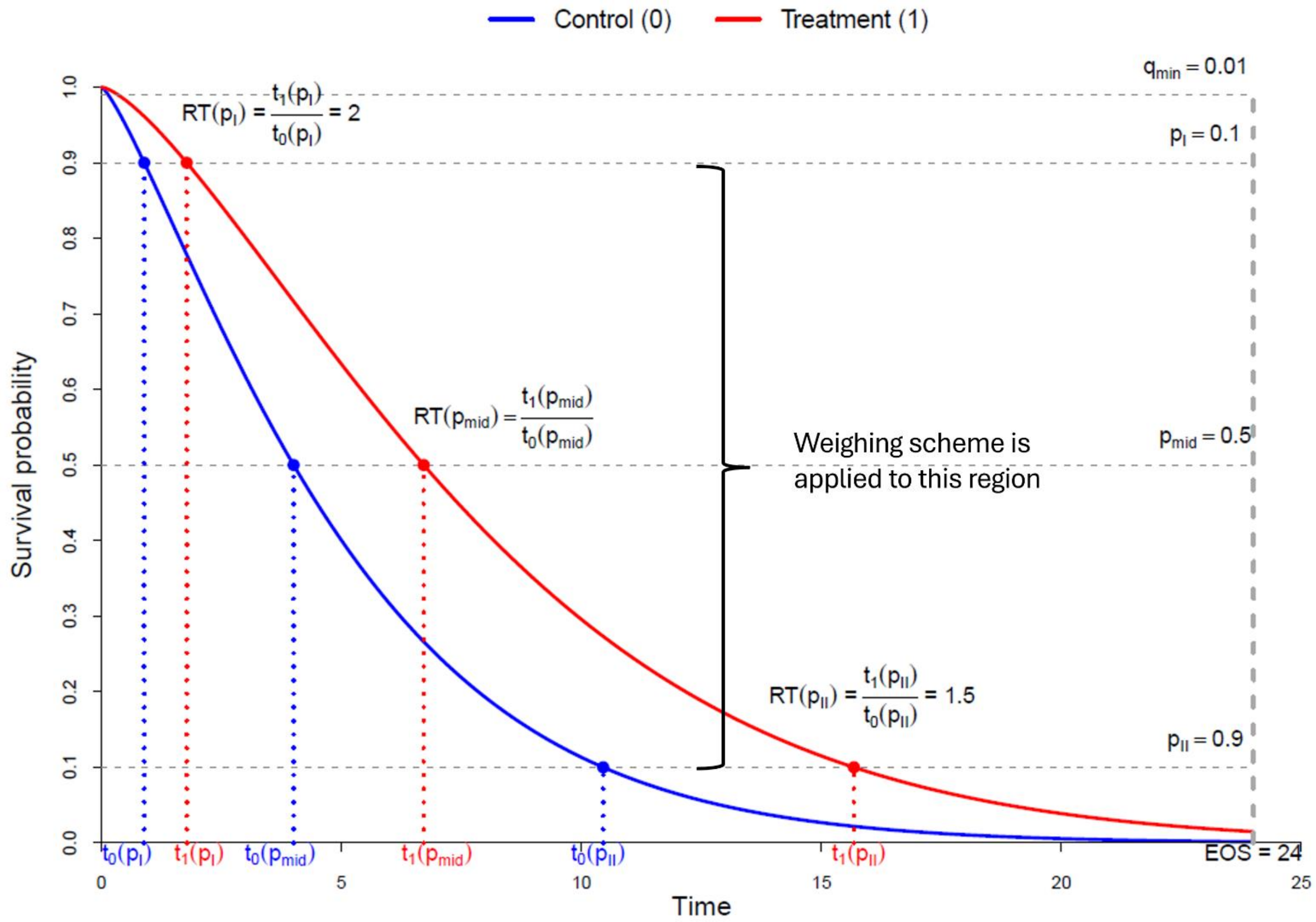

Figure 3. Weibull hazards as a function of time for the control and treatment arm

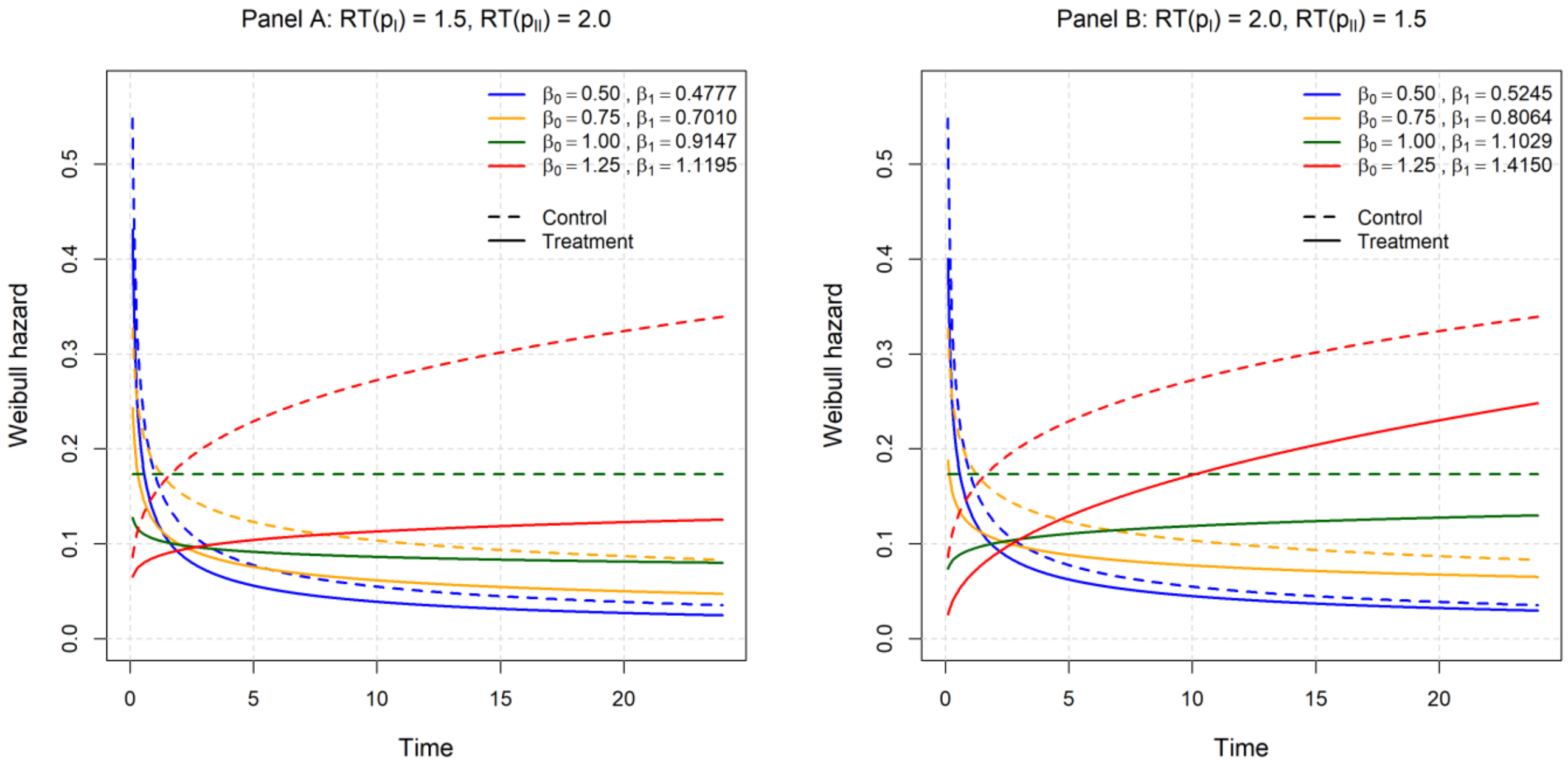

Figure 4. Sensitivity analysis of sample sizes for different weighting schemes

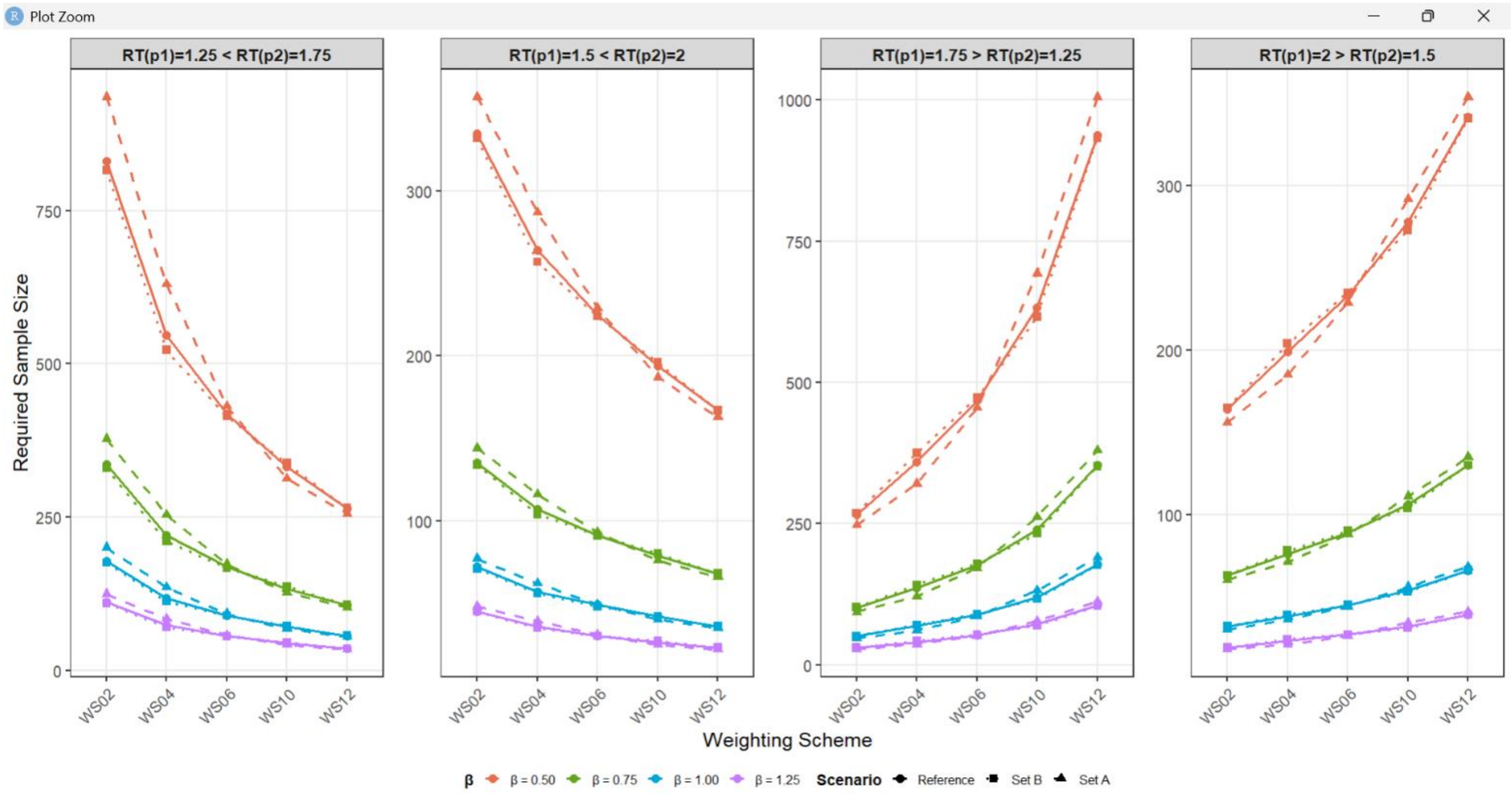